\documentclass[cits]{PoS}

\usepackage{amsmath}
\usepackage{psfrag}

\title{Electroweak penguins in isospin-violating $B_s$ decays}

\ShortTitle{Electroweak penguins in isospin-violating $B_s$ decays}

\author{Lars Hofer$^{ab}$, \speaker{Dominik Scherer}$^{a}$  and Leonardo Vernazza$^c$ \thanks{TTP10-51, MZ-TH/10-46, SFB/CPP-10-129. Supported by Deutsche Forschungsgemeinschaft SFB/TR9, BMBF contract 05H09WWE, Cusanuswerk, Studienwerk Villigst and Alexander-von-Humboldt foundation} \\
         \llap{$^a$} Institut f\"ur Theoretische Teilchenphysik,\\
Karlsruhe Institute of Technology, D--76128 Karlsruhe, Germany\\
 \llap{$^b$} Institut f\"ur Theoretische Physik und Astrophysik, \\
Universit\"at W\"urzburg, D--97074 W\"urzburg, Germany\\
  \llap{$^c$} Institut f\"ur Physik (THEP),\\ Johannes Gutenberg-Universit\"at, D--55099 Mainz, Germany\\
         E-mail: \email{lars.hofer@physik.uni-wuerzburg.de}, \email{dominik@particle.uni-karlsruhe.de}, \email{vernazza@uni-mainz.de}}

\abstract{The $2.5\sigma$ discrepancy between theory and experiment observed in the difference $\Delta A_\text{CP}=$ \linebreak\mbox{$A_\text{CP}(B^-\to\pi^0 K^-)-A_\text{CP}(\bar{B}^0\to\pi^+ K^-)$} can
be explained by a new electroweak (EW) penguin amplitude. Motivated by this result, we have analyzed 
the purely isospin-violating decays $\bar B_s\to\phi\pi^0$ and $\bar B_s\to\phi\rho^0$, which are dominated by EW penguins. Our results extend the analysis in \cite{ourproceeding} and have recently been published in \cite{ourpaper}. Here we give a brief overview of the outcome.

We show that in presence of a new EW penguin amplitude the two $B_s$ branching ratios can be
enhanced by an order of magnitude without violating any constraints from other hadronic $B$ decays. This makes them very interesting modes for LHCb and Super $B$ factories. We perform both a model-independent analysis and a study within realistic New Physics (NP) models such as a modified-$Z^0$-penguin scenario, 
a model with an additional $Z^{\prime}$ boson and the MSSM, including a fit to $B\rightarrow\pi K$ data and the relevant experimental constraints throughout. In the model-independent case we study effective $b\rightarrow s \bar q q$ couplings and distinguish between several possible chirality structures. Constraints arise from a large number of hadronic $B$ decays such as $B\rightarrow \pi K^{(*)}, \rho K^{(*)},\phi K^{(*)}$ etc. The preferred fit regions are rather large and allow for order-of-magnitude enhancements (see plots). In concrete models the new amplitude can often be correlated with other flavour phenomena, such as semileptonic $B$ decays and $B_s$-$\bar{B}_s$ mixing, which set stringent constraints on the enhancement of the two $B_s$ decays. In particular we find that, contrary to claims in the literature, EW penguins in the MSSM can reduce the discrepancy in $\Delta A_\text{CP}$ only marginally. Consequently no visible enhancement of $\bar B_s\to\phi\pi^0,\phi\rho^0$ is expected for this model. As byproducts of our work we update the Standard Model (SM) predictions  to $BR(\bar B_s\to\phi\pi^0)=1.6^{+1.1}_{-0.3}\cdot 10^{-7}$ and $BR(\bar B_s\to\phi\rho^0)=4.4^{+2.7}_{-0.7}\cdot 10^{-7}$ and perform a state-of-the-art analysis of $B\rightarrow \pi K$ amplitudes in QCD factorisation (QCDF).
}

\FullConference{35th International Conference of High Energy Physics\\
		 July 22-28, 2010\\
		 Paris, France}

\begin{document}

\paragraph{Amplitude structure of $\bar B_s \rightarrow \phi\pi^0$ and $\bar B_s \rightarrow \phi\rho^0$:}
The two decay modes discussed here are \emph{pure} $\Delta I=1$ transitions, which means that there are no QCD penguin contributions at all. Calculating both amplitudes in QCDF the dominant SM contribution comes from a $Z^0$ penguin with pollution from a CKM- and colour-suppressed tree amplitude and possibly from OZI-suppressed singlet annihilation. We provide convenient approximate expressions for the amplitudes in \cite{ourpaper}. 
\paragraph{Effects of new EW penguin amplitudes:}
We study the effects of new EW penguin amplitudes in different ways: \\[-0.7cm]
\begin{itemize}\addtolength{\itemsep}{-0.52\baselineskip}
 \item[a)] We parameterise the new amplitudes in a model-independent way via complex numbers $q_i$. Here we show the enhancement factors (circles) of $BR(\bar B_s \rightarrow \phi\pi^0)$ (left) and $BR(\bar B_s \rightarrow \phi\rho^0)$ (right) w.r.t. the SM as a function of $q_9$, corresponding to an EW $b_L\rightarrow s_L\bar q_L q_L$ penguin. For $|q_9|=1$ the corresponding SM penguin receives a $100\%$ correction. Dark green areas are allowed by constraints from hadronic $B$ decays while the solid black lines mark the preferred ($1\sigma$) region of the $B\rightarrow\pi K$ fit. Lighter areas and lines only take into account observables particularly sensitive to isospin-violation. The hatched rings mark the theoretical uncertainty of the SM branching ratios. We see that an order-of-magnitude enhancement is possible for both branching ratios. The same is true for several right-handed and mixed NP amplitudes.
\begin{center} 
 \psfrag{Imq}[tc][tc][1][0]{\small{$\text{Im}(q_9)$}}
\psfrag{Req}[Bc][Bc][1][0]{\small{$\text{Re}(q_9)$}}
\includegraphics[width=0.67\textwidth]{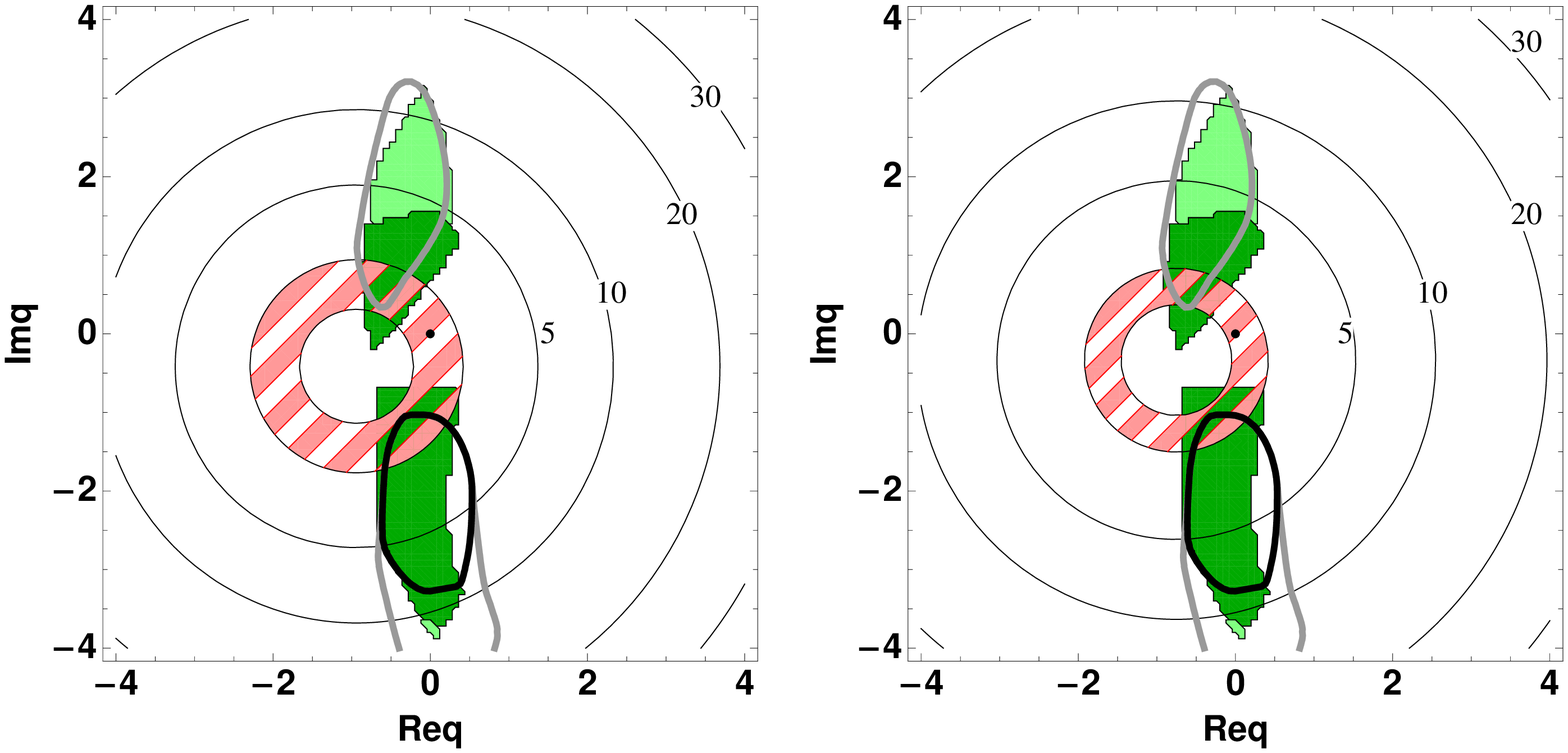}
\end{center}
 \item[b)] We calculate the new amplitudes in concrete NP models. We find that flavour-changing Z-couplings cannot produce large effects because of tight constraints from semileptonic $B$ decays. These constraints can be relaxed in $Z'$ models since the $Z'$ couplings to leptons are unknown. The most important constraint in this model comes from $B_s-\bar B_s$ mixing, it allows for enhancements of up to a factor of $\sim 5$ in the $B_s$ decays. In the MSSM we find that no large isospin-violating effects are possible at all, neither in $\Delta A_\text{CP}$ nor in     
the two $B_s$ decays.  
\end{itemize}
\paragraph{Conclusion:} We strongly encourage experimental efforts towards a measurement of $\bar B_s \rightarrow \phi\pi^0$ and $\bar B_s \rightarrow\phi\rho^0$ at LHCb and Super $B$ factories.

\end{document}